\documentclass[letterpaper, 10 pt, conference]{ieeeconf}  
\usepackage{graphicx}
\usepackage{stfloats}
\usepackage{subcaption}
\usepackage{tikz}
\usepackage{breqn}
\usepackage{amsmath}

\IEEEoverridecommandlockouts                              

\title{\LARGE \bf
Multimodal Sensing and Machine Learning to Compare Printed and Verbal Assembly Instructions Delivered by a Social Robot}

\author{Ruchik Mishra, Laksita Prasanna, Adair Adair, and Dan O. Popa$^{*}$
\thanks{*This project was supported in part by the National Institutes of Health (NIH) and the National Science Foundation (NSF) through grants \#1838808 and \#1849213. Authors are with the Louisville Automation and Robotics Research Institute (LARRI), University of Louisville, Kentucky, USA. Contact email {\tt\small ruchik.mishra@louisville.edu}.}
}

\begin{document}

\maketitle
\thispagestyle{empty}
\pagestyle{empty}


\begin{abstract}
In this paper, we compare a manual assembly task communicated to workers using both printed and robot-delivered instructions. The comparison was made using physiological signals (blood volume pulse (BVP) and electrodermal activity (EDA)) collected from individuals during an experimental study. In addition, we also collected responses of individuals using the NASA Task Load Index (TLX) survey. Furthermore, we mapped the collected physiological signals to the responses of participants for NASA TLX to predict their workload. For both the classification problems, we compare the performance of Convolutional Neural Networks (CNNs) and Long-Short-Term Memory (LSTM) models. Results show that for our CNN-based approach using multimodal data (both BVP and EDA) gave better results than using just BVP (approx. 8.38\% more) and EDA (approx 20.49\% more). Our LSTM-based model too had better results when we used multimodal data (approx 8.38\% more than just BVP and 6.70\% more than just EDA). Overall, CNNs performed better than LSTMs for classifying physiologies for paper vs robot-based instruction by 7.72\%. The CNN-based model was able to give better classification results (approximately 17.83\% more on an average across all responses of the NASA TLX) within a few minutes of training compared to the LSTM-based models.

\end{abstract}

\begin{figure*}[!b]
    \centering
    \includegraphics[width=0.72\textwidth]{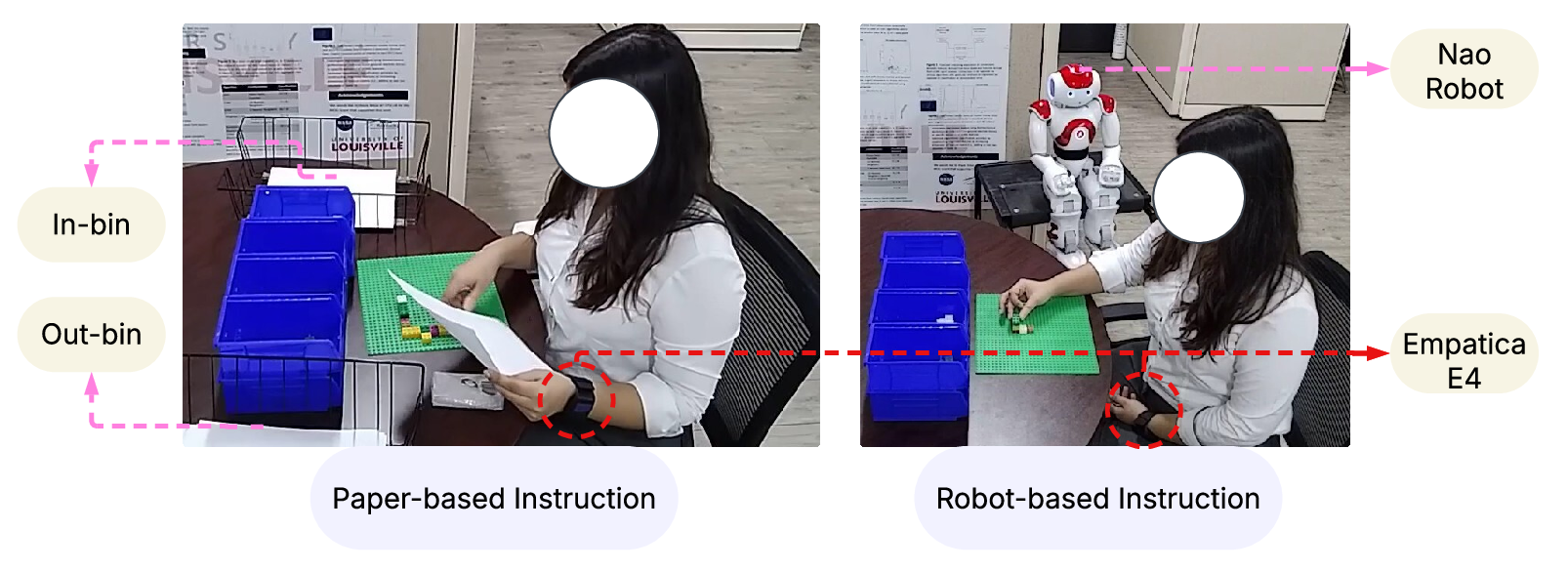}
    \caption{Printed vs Robot-based instruction for a simulated assembly task.}
    \label{fig:Teaser_image}
\end{figure*}
\section{Introduction}
The use of robots for industrial assembly is increasing \cite{llale2020review, sherwani2020collaborative}. In the last decade, we have seen extraordinary growth and development of robotic systems to pursue tasks that involve human robot interaction \cite{zhang2022reinforcement, liu2021deep,borboni2023expanding}. These tasks not only include simple pick-and-place tasks but also extend to a wide range of highly dexterous maneuvers \cite{chen2014hand,bonivento1989control}.
Even beyond dexterity of robotic manipulators, the application of robotics extends to more meaningful human-robot interaction situations like in an industrial assembly task \cite{quitter2017humanoid} using a humanoid robot. These humanoid robots coexit in these industrial settings in different forms: cobots, or collaborative robots where they are physically involved with humans and share a workspace \cite{cohen2019strategic, akella1999cobots}. With the advent of Artificial Intelligence, the application of robots in industrial settings has become even more sophisticated. Using sensor data to better understand human perceptions of the robot and the robot's perception of the human's workload is an essential part of HRI of shared workspaces \cite{aygun2022investigating}.  

Robots as instructors for assembly tasks have been explored with various perspectives. The authors in \cite{quitter2017humanoid} use a humanoid robot, Baxter, for training workers to familiarize themselves with the assembly of a gearbox. Their work mainly focused on the performance of the individuals and the perception of their participants using the Godspeed survey. However, they did not compare their approach of using robots to the conventional paper-based, or other modes of instruction.

The authors in \cite{blattgerste2017comparing} compared three modes of instruction for an assembly task: Augmented Reality (AR), conventional pictorial instruction using smartphones, Microsoft HoloLens and Epson Moverio BT-200, and a paper-based instruction mode. The authors collected participants' responses on the NASA TLX, but do not present any intrinsic modality to evaluate or predict cognitive workload except for subjective measures. The authors proposed another augmented reality-based instruction in \cite{funk2015benchmark}. But similar to the authors in \cite{blattgerste2017comparing}, they only used subjective measures to compare AR and paper-based instructions. All participants performed simulated assembly tasks.

In this paper, we monitor human physiological responses and other participant-reported subjective measures to compare printed and social robot-delivered assembly instructions. We collect BVP and EDA signals using an Empatica E4 wrist band during the manual completion of the task using printed or robot-spoken assembly instructions. Additionally, we map these physiological signals to the participant survey responses using the NASA TLX \cite{hart2006nasa}. We used Lego brick assemblies to mimic similar industrial operations, as evidenced in their prior similar uses \cite{blattgerste2017comparing,jeffri2020problems}. We collected and analyzed survey and physiological data from 22 participants in our study. The research contributions of the paper are as follows:
\begin{itemize}
    \item We evaluated the physiological differences between printed and robot-based instructions data for teaching simulated industrial assembly tasks using Empatica E4 wrist band.
    \item We used machine learning and compared CNN and LSTM-based models for the collected wearable data. We also predicted the scores (formulated as a classification problem) of the participants using our CNN and LSTM-based approaches using the collected physiological measures
    \item We found our CNN-based models to perform well for both classification problems. In addition, for our second classification problem, CNNs outperformed LSTMs by training on the first three minutes of training data. 
\end{itemize}
Our findings indicate physiological differences in participants between printed and robot-delivered instructions. We could also predict the participants' scores on the NASA TLX using these physiological measures. This indicates that the use of collected BVP and EDA signals through a wearable can replace the subjective surveys used to evaluate the cognitive workload of participants. We also analyze how different numbers of training samples (represented as how early in the session) can our deep learning models start to generalize the test data well (through classification accuracy).

This paper has been organized in the following manner: Section \ref{methodology} describes the methodology;  Section \ref{problem_form} describes the problem formulation; Section \ref{results_and_disc} discusses the results we obtained. Finally, Section \ref{conclusion} describes the conclusion.
\begin{figure}[h!]
    \centering
    \includegraphics[width=0.8\linewidth]{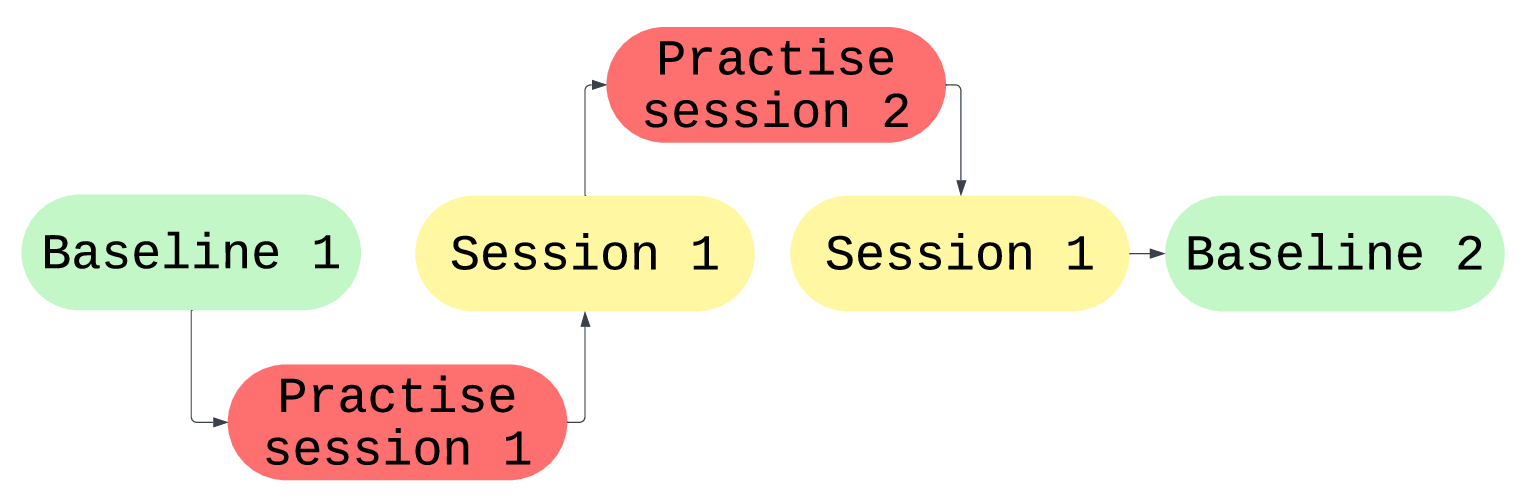}
    \caption{Study design}
    \label{fig:pipeline}
\end{figure}
\section{Methodology}\label{methodology}
\subsection{Data Acquisition}\label{data_acq}
We recruited 22 participants for this study. All the participants were between 18 and 30 years of age. We record BVP and EDA from each participant using the Empatica E4 wearable sensor. Informed consent was taken from all the participants. They also had the option to leave the session at any point in time. This study was approved by the Institutional Review Board (IRB) \# 18.0726.  

\subsection{Study Design}\label{study_design}
\begin{figure*}[!h]
  \centering
  \begin{subfigure}{0.7\textwidth}
    \centering
    \includegraphics[width=\linewidth]{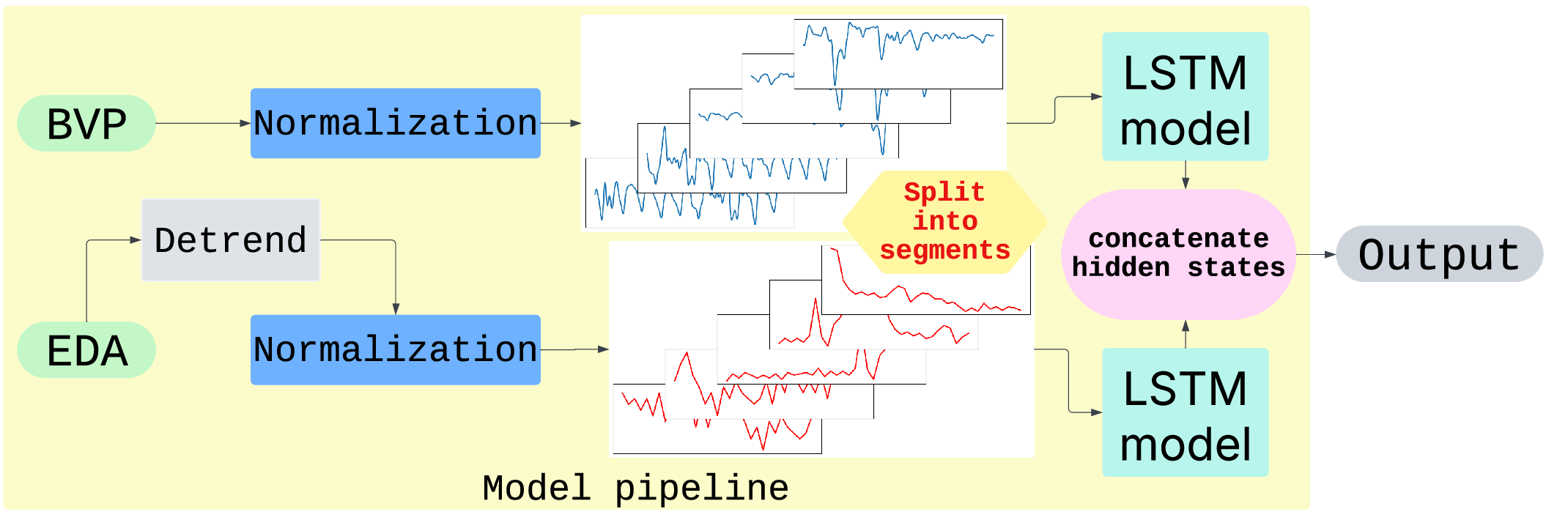} 
    \caption{Mutimodal pipeline using LSTM}
    \label{fig:LSTM_pipeline}
  \end{subfigure}
  \hfill
  \begin{subfigure}{0.75\textwidth}
    \centering
    \includegraphics[width=\linewidth]{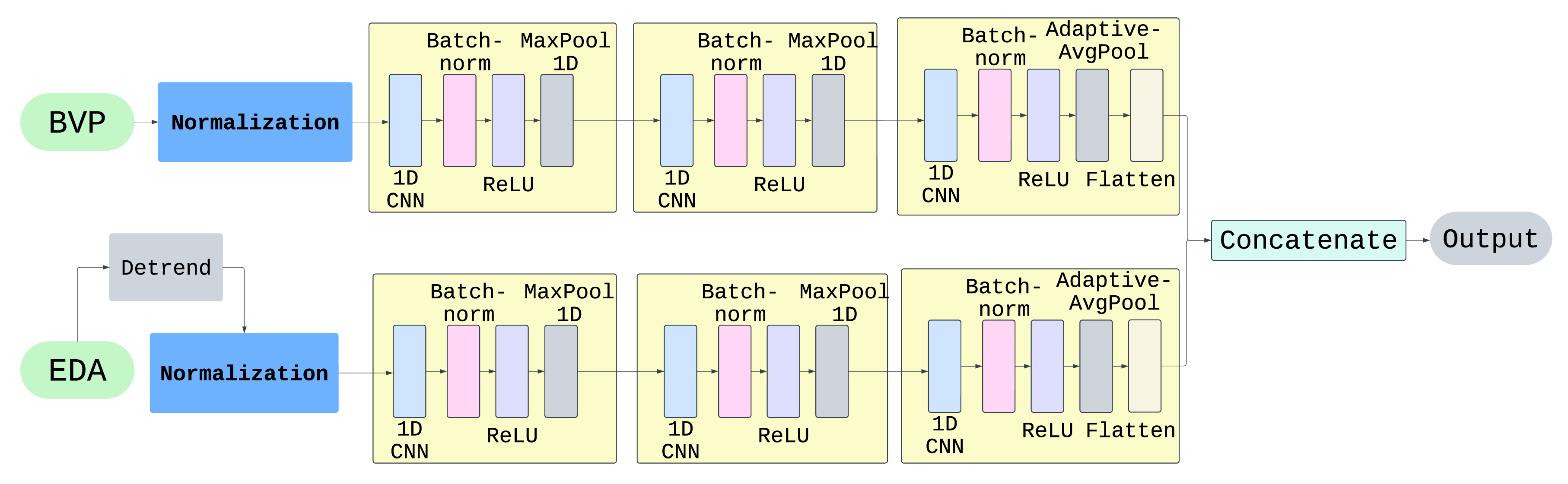} 
    \caption{Multimodal pipeline using CNN}
    \label{fig:CNN_pipeline}
  \end{subfigure}
  \caption{Multimodal pipeline to combine BVP and EDA data for both classification tasks: 1) classifying physiologies for printed vs robot-based instruction task, and 2) predicting the responses of NASA TLX based on physiologies.}
  \label{fig:model_pipelines}
\end{figure*}
Each participant in the study goes through several segments, as illustrated in Figure \ref{fig:pipeline}. These include an initial baseline session, a robot trial, a robot actual run, a paper trial, a paper actual run, and a second baseline session. During each three—minute baseline, participants sit alone in an uncluttered lab space facing a camera that records their facial expressions, and wear an Empatica E4 to capture physiological data.

For the trial sessions of both the robot and paper based instructions, participants complete a straightforward 4-step task to familiarize them with the Lego assembly process and terminology used in the robot’s and printed instructions. During these trials, participants are encouraged to ask the researchers as many questions as needed, and their performance is not evaluated.

Participants receive 16-step instructions to complete the Lego assembly in the actual runs for both conditions. During these sessions, researchers leave the room so each participant works independently. For the printed instruction task, participants view one instruction at a time. As shown in Figure \ref{fig:Teaser_image}, the printed instructions are arranged sequentially in the ‘In-bin’ (Steps 1–16), with each paper containing only one step. After reading an instruction and placing the corresponding Lego piece, the instruction is moved to the ‘Out-bin.’ If a participant wishes to review a previous step, they must transfer papers one by one from the ‘Out-bin’ back to the ‘In-bin’ until they locate the desired instruction, ensuring that only a single instruction is visible at any moment.

For the robot task, participants verbally gave the instruction number to the robot to receive the corresponding instruction for a step. They can also request a repetition or return to an earlier step, similar to the paper-based method. The robot offers positive reinforcement by giving a fist bump after the eighth instruction, and participants in the paper task are similarly prompted to provide themselves with a fist bump after eight steps.

In addition to recording physiological responses during these tasks, participants completed the NASA Task Load Index \cite{hart2006nasa} after the actual runs for both the printed and robot instruction task. This provided a means to assess and compare the workload imposed by each method.

\subsection{Pre-processing}\label{preprocessing}
Since we record two baselines (in the beginning and at the end of the study), we check for baseline drift using a paired t-test in python (discussed more in Section \ref{paper_vs_robot_physio}). In case of baseline drift, we perform de-trending of the signal and check for differences in the two baselines again \cite{posada2020innovations, info:doi/10.2196/72093}. 
\section{Problem Formulation}\label{problem_form}
In this paper, we propose solutions to two classification problems: 1) classifying physiological signals between printed and robot-based instruction tasks using binary classification, and 2) predicting participants' responses on the NASA TLX using physiological signals. For both of these models, we compare our CNN and LSTM-based pipelines.  
For each participant, $p \in \left\{\textrm{P1}, \dots, \textrm{P22}\right\}$, the BVP, denoted by $\textbf{x}_{p, BVP, k}$ and the EDA signal, denoted by ${x}_{p, EDA, k}$ is split into smaller segments of $n$ seconds, where $k \in \left\{\ \textrm{Baseline 1}, \textrm{Session 1},\textrm{Session 2},\textrm{Baseline 2}\right\}$ for sessions $l \in \left\{\ \textrm{Session 1},\textrm{Session 2}\right\}$. This normalization for the BVP signal is done with respect to the first baseline as:
\begin{equation}\label{eq:bvp_norm}
    \textbf{x}_{p,BVP,l,norm} = \textbf{x}_{p,BVP,l} - \textbf{x}_{p,BVP,\textrm{Baseline 1}}  
\end{equation}
\begin{equation}\label{eq:split_bvp}
    f_{split}( \textbf{x}_{p,BVP,l}) = \left[\textbf{\textit{x}}_{1,BVP,l}, \dots, \textbf{\textit{x}}_{N,BVP,l,norm}\right]
\end{equation}
where $N = \frac{len(\textbf{x}_{p,BVP,l,norm})}{n\times sr_{BVP}}$, $sr_{BVP}$ is the sampling rate of the BVP signal. For the EDA signal, we first detrend all parts of the signal (Baseline 1, Session 1, Session 2, and Baseline 2), then apply the normalization process:
\begin{equation}
    \textbf{x}_{p,EDA,k, detrend} = detrend(\textbf{x}_{p,EDA,k, detrend}) 
\end{equation}
\vspace{-0.6cm}
\begin{dmath}
        \textbf{x}_{p,EDA,l,norm}    = \textbf{x}_{p,EDA,l, detrend} - \\detrend(\textbf{x}_{p,EDA,Baseline1, detrend})
\end{dmath}
\begin{equation}\label{eq:split_eda}
    f_{split}( \textbf{x}_{p,EDA,l,norm}) = \left[\textbf{\textit{x}}_{1, EDA,l}, \dots, \textbf{\textit{x}}_{M, EDA,l}\right]
\end{equation}
where, $M = \frac{len(\textbf{x}_{p,EDA,l,norm})}{n\times sr_{EDA}}$, and $sr_{EDA}$ is the sampling rate of the EDA signal. 
\subsection{Unimodal classification for printed vs robot-based instruction}
For the LSTM model, we use equation \ref{eq:unimodal_lstm}:
\begin{equation}\label{eq:unimodal_lstm}
    \textbf{h}_{t,j}, \textbf{c}_{t,j} = LSTMCell(\textbf{\textit{x}}_{t,i,j,l}^{\textrm{\textbf{T}}},\textbf{h}_{t-1,i,j},\textbf{c}_{t-1,i,j})
\end{equation}
\begin{equation}\label{eq:unimodal_forward}
    \textbf{z}_{unimodal} = W_{unimodal}\textbf{h}_{t} + \textbf{b}_{unimodal}
\end{equation}
where $LSTMCell(.)$ is the typical LSTM cell as described in \cite{yu2019review}, $i$ is the segment number (described in equation \ref{eq:split_bvp} and \ref{eq:split_eda}), $j$ is either BVP or EDA, $W_{unimodal} \in \mathcal{R}^{c \times H_{j}}$, $H_{j}$ is the hidden dimension of modality $j$, and $\textbf{b}_{unimodal} \in \mathcal{R}^{c}$. Here since it is a binary classification, $c = 2$.

Similarly, the CNN pipeline is given by:
\begin{subequations}
\begin{align}
        \textbf{z}_{CNN,j}^{(1)} = ReLU\left(\textrm{BN}\left(Conv_{1}(\textbf{\textit{x}}_{j}^{\textrm{\textbf{T}}})\right)\right)
        \end{align}
        \vspace{-0.6cm}
        \begin{align}
            \textrm{\textbf{p}}_{j}^{(1)} =  \textrm{Pool}( \textbf{z}_{CNN,j}^{(1)})
        \end{align}
        \vspace{-0.6cm}
        \begin{align}
              \textbf{z}_{CNN,j}^{(2)} = ReLU\left(\textrm{BN}\left(Conv_{2}(\textrm{\textbf{p}}_{j}^{(1)})\right)\right)
        \end{align}
        \vspace{-0.6cm}
        \begin{align}
            \textrm{\textbf{p}}_{j}^{(2)} =  \textrm{Pool}( \textbf{z}_{CNN,j}^{(2)})
        \end{align}
        \vspace{-0.6cm}
        \begin{align}
            \textbf{z}_{CNN,j}^{(3)} = ReLU\left(\textrm{BN}\left(Conv_{3}(\textrm{\textbf{p}}_{j}^{(2)})\right)\right)
        \end{align}
        \vspace{-0.6cm}
        \begin{align}
            \textrm{\textbf{g}}_{CNN,j} = \textrm{Flatten}\left(\textrm{GAP}\left(\textbf{z}_{CNN,j}^{(3)}\right)\right)
        \end{align}
        \vspace{-0.6cm}
        \begin{align}
            \textrm{\textbf{y}}_{CNN,j} = W_{CNN,u,j}\textrm{\textbf{g}}_{CNN,j} + \textrm{\textbf{b}}_{CNN,u,j}
        \end{align}
where, $Conv_{l}$ represents the $l^{th}$ convolution layer with stride one and padding 1, BN represents batch normalization, GAP represents Global Average Pooling, $W_{CNN,u,j} \in \mathcal{R}^{2 \times 64}$, and $\textrm{\textbf{b}}_{CNN,u,j} \in \mathcal{R}^{2\times 1}$.  
\end{subequations}
\subsection{Multimodal data classification into paper vs robot-based task}
For multimodal data, we concatenate the hidden states of the LSTM pipelines (each for BVP and EDA) and then use a linear layer for the classification. We calculate $h_{fusion} \in \mathcal{R}^{1\times H_{BVP} + H_{EDA}}$
\begin{equation}\label{eq:multimodal_lstm}
    h_{fusion} = \left[h_{T_{BVP},BVP} \hspace{0.2cm} h_{T_{EDA},EDA}\right]^{\textrm{\textbf{T}}}
\end{equation}
where $T_{BVP}$ and $T_{EDA}$ are the sequence lengths of the BVP and EDA signal segments respectively.
So, equation \ref{eq:unimodal_forward} now becomes:
\begin{equation}\label{eq:multimodal_forward_pass_robot_vs_paper}
      \textbf{z}_{multimodal} = W_{multimodal}\textbf{h}_{fusion} + \textbf{b}_{multimodal}
\end{equation}
where the symbols have their usual meaning as described in equation \ref{eq:unimodal_lstm}.
For our CNN based pipeline, equation \ref{eq:CNN_fusion} described how we concatenate features after the Global Average Pooling Layer:
\begin{subequations}\label{eq:CNN_fusion}
    \begin{align}
        \textrm{\textbf{g}}_{CNN,fusion} = \left[\textrm{\textbf{g}}_{CNN,BVP} \hspace{0.2cm} \textrm{\textbf{g}}_{CNN,EDA}\right]
    \end{align}
    \vspace{-0.6cm}
    \begin{align}
        \textrm{\textbf{y}}_{CNN,fusion} = W_{CNN,m}\textrm{\textbf{g}}_{CNN,fusion} + \textrm{\textbf{b}}_{CNN,m,fusion}
    \end{align}
\end{subequations}
where $W_{CNN,m} \in \mathcal{R}^{2 \times 128}$ and $\textrm{\textbf{b}}_{CNN,m,fusion} \in \mathcal{R}^{2\times 1}$.
\subsection{Multimodal data for classifying workload}
We use the physiological signals (BVP and EDA) recorded from the participants to predict their response on the NASA TLX. We formulate this as a multi-output classification problem. We use the same $LSTMCell(.)$ as used in equation \ref{eq:unimodal_lstm}, then we change the forward pass as in equation \ref{eq:forward_pass_TLX}:
\begin{equation}\label{eq:forward_pass_TLX}
    \textbf{z}_{k} = W_{k} \textbf{h}_{fusion} + \textbf{b}_{k}
\end{equation}
where $k \in K = \left\{\textrm{MD},\textrm{TD},\textrm{PD},\textrm{Eff},\textrm{Frus}\right\}$, $W_{k} \in \mathcal{R}^{|\mathcal{K}| \times H_{BVP} + H_{EDA}}$ and $\textbf{b}_{k} \in \mathcal{R}^{|\mathcal{K}|}$. Then we apply softmax for converting the logits into class probabilities as:
\begin{equation}\label{eq:softmax_logits_TLX}
 \widehat{y}_{k} = \textrm{softmax}(\textbf{z}_{k})   
\end{equation}
During training, we add the losses of each of these outputs as:
\begin{equation}
    \mathcal{L}_{\textrm{total}} = \frac{1}{|K|}\sum_{k}\mathcal{L}_{k}
\end{equation}
where $\mathcal{L}_{k}$ is the cross-entropy loss.
Similarly, for the CNN pipeline, we concatenate the features after flattening the $\textrm{GAP}(.)$:
\begin{subequations}
    \begin{align}\label{eq:CNN_nasa_y}
        y_{CNN,k,fusion} = W_{k,CNN}\textrm{\textbf{g}}_{CNN,fusion} + \textbf{b}_{k,fusion}
    \end{align}
    \vspace{-0.6cm}
    \begin{align}
        \widehat{y}_{k,CNN} = \textrm{softmax}(y_{CNN,k,fusion})
    \end{align}
    \vspace{-0.6cm}
    \begin{align}
        \mathcal{L}_{total,CNN} = \frac{1}{|K|}\sum_{k}\mathcal{L}_{k,CNN}
    \end{align}
\end{subequations}
where $W_{k,CNN} \in \mathcal{R}^{|\mathcal{K}| \times 128}$ and $\textbf{b}_{k,fusion} \in \mathcal{R}^{|\mathcal{K}| \times 1}$.
\vspace{-0.3cm}
\begin{figure}[h!]
    \centering
    \includegraphics[width=0.85\linewidth]{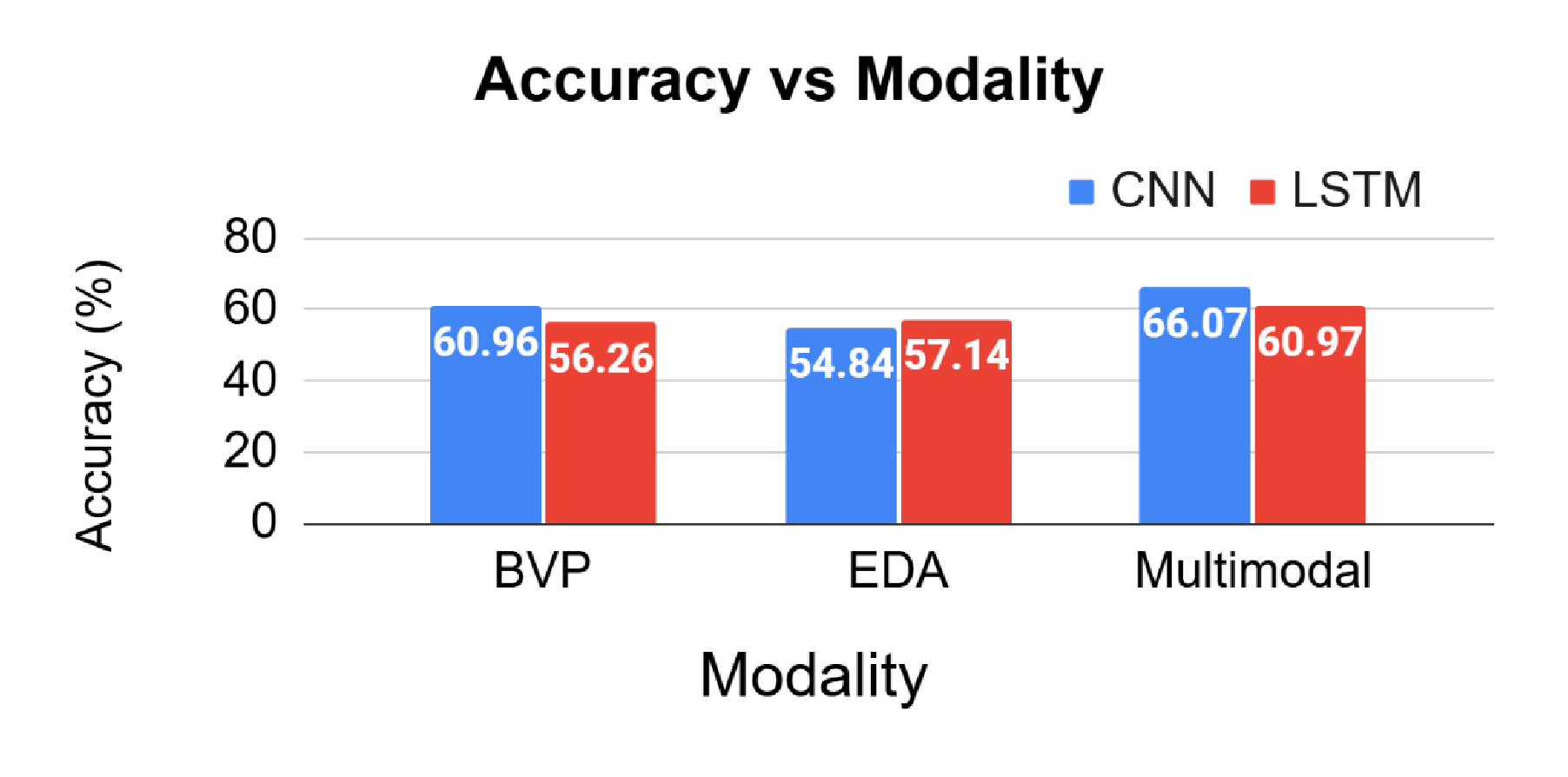}
    \caption{Comparing CNN and LSTM models' percentage accuracy for unimodal vs multimodal use of physiological signals for classifying paper vs robot-based instruction.}
    \label{fig:acc_vs_modal}
\end{figure}
\vspace{-0.6cm}
\begin{figure}[h!]
    \centering
    \includegraphics[width=1.0\linewidth]{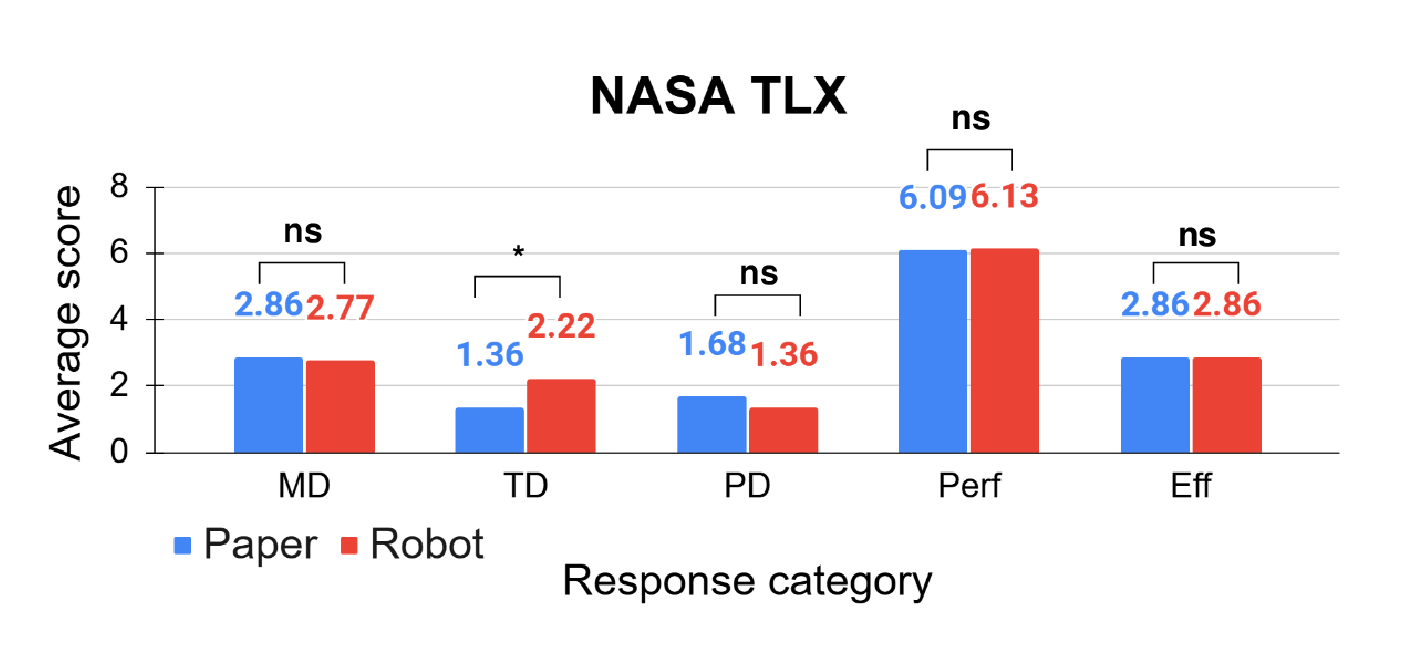}
    \caption{Paired t-test results on NASA TLX for comparing paper vs robot-based instructions for MD, TD, PD, Perf, and Eff. ns: no significance and $(*)$ denotes $p-$value $<0.05$}
    \label{fig:p_val_figure}
\end{figure}
\vspace{-0.4cm}
\begin{figure}[h!]
    \centering
    \includegraphics[width=0.7\linewidth]{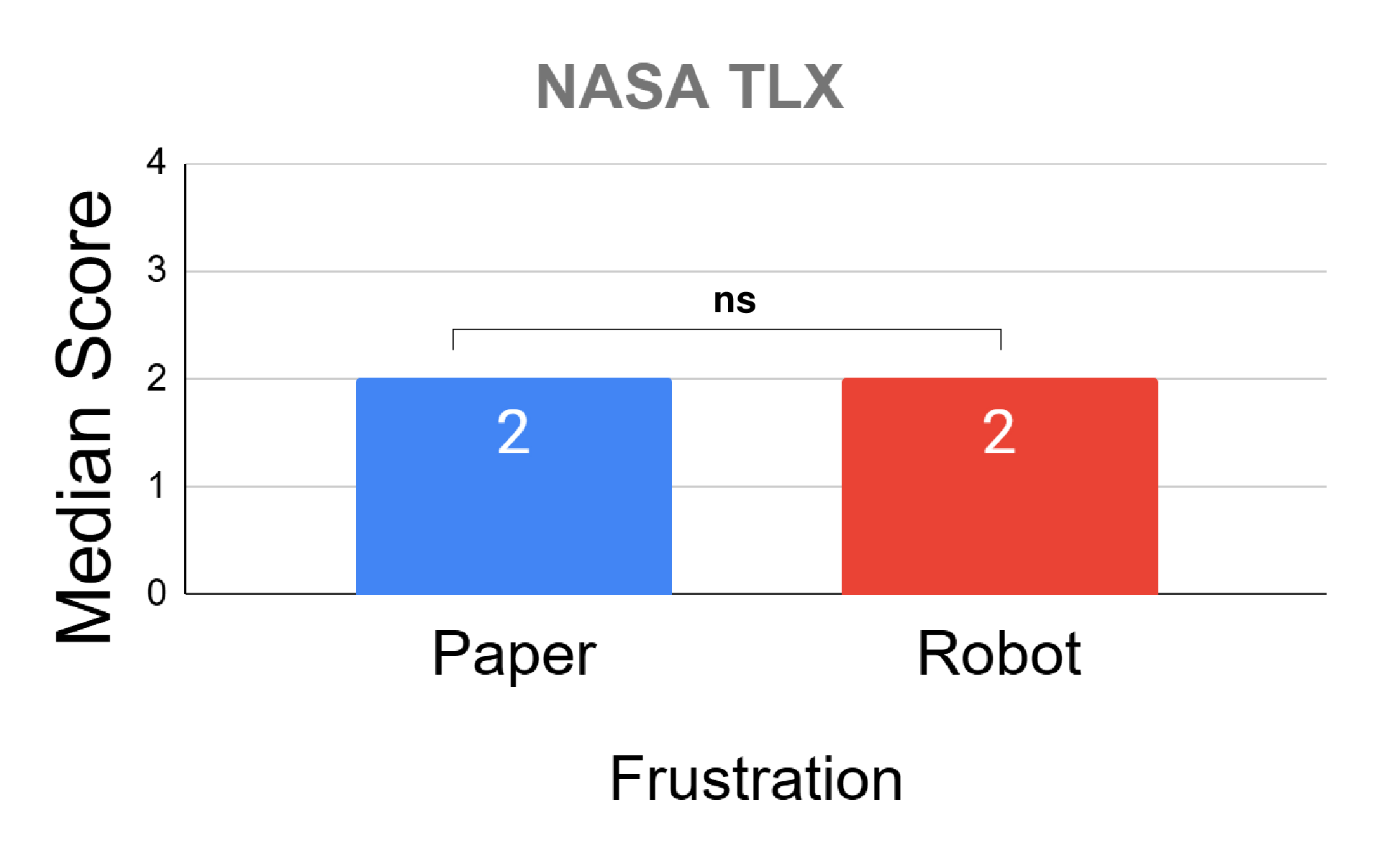}
    \caption{Wilcoxon signed rank test on NASA TLX for comparing paper vs robot-based instructions for Frustration. }
    \label{fig:wilcoxon}
\end{figure}
\begin{figure*}[!h]
  \centering
  \begin{subfigure}{0.30\textwidth}
    \centering
    \includegraphics[width=\linewidth]{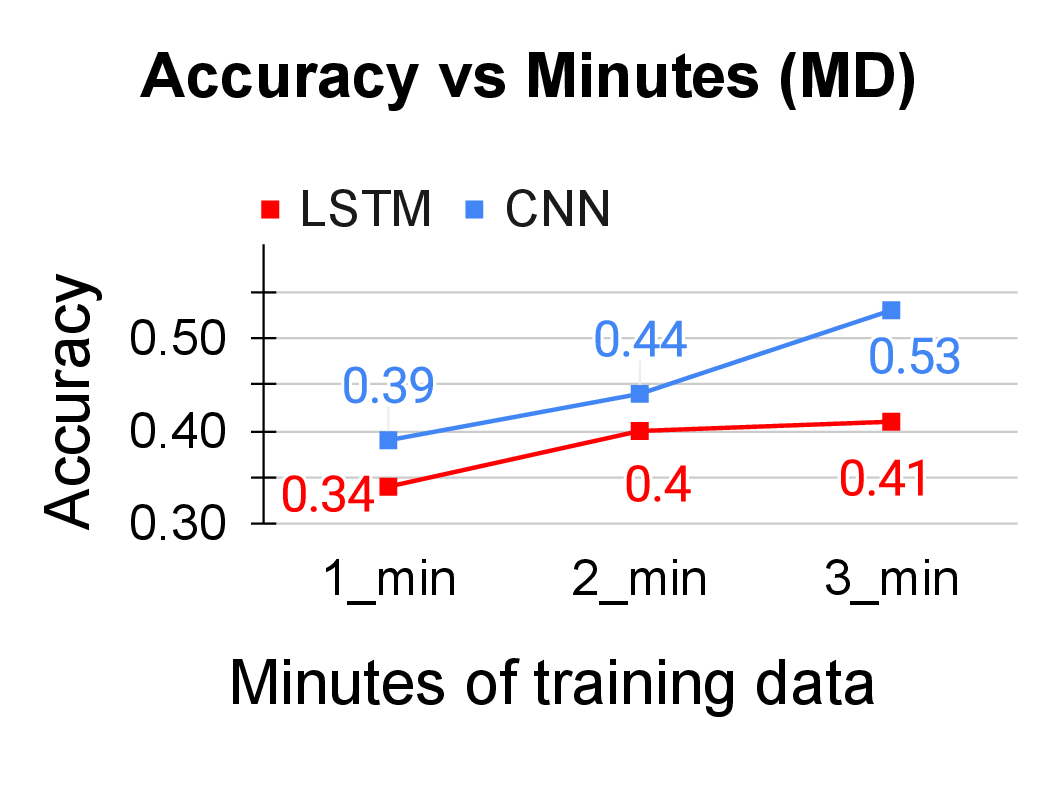} 
    \caption{CNN vs LSTM for MD}
    \label{fig:CNN_LSTM_MD}
  \end{subfigure}
  \hfill
  \begin{subfigure}{0.30\textwidth}
    \centering
    \includegraphics[width=\linewidth]{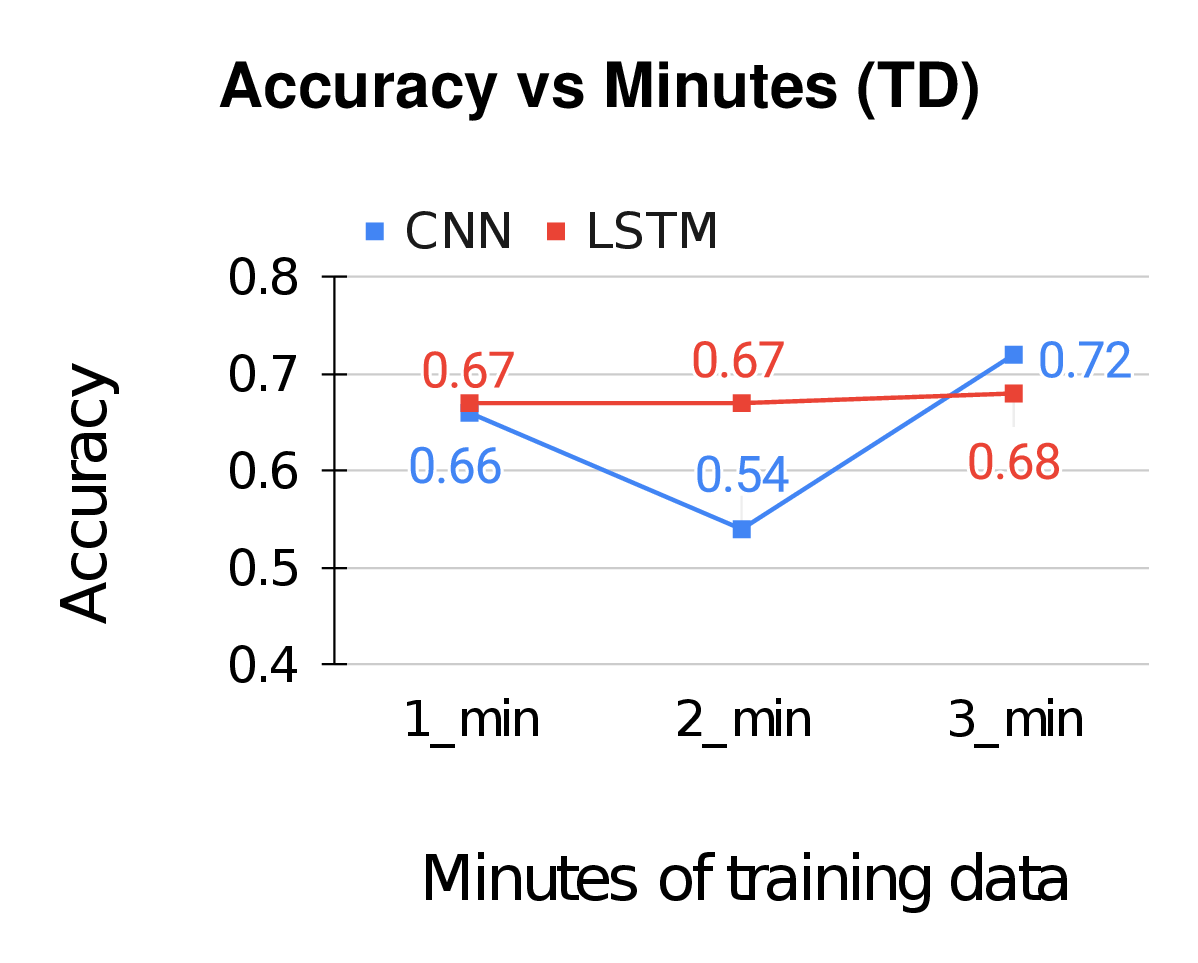} 
    \caption{CNN vs LSTM for TD}
    \label{fig:CNN_LSTM_TD}
    \hfill
  \end{subfigure}
  \hfill
   \begin{subfigure}{0.32\textwidth}
    \centering
    \includegraphics[width=\linewidth]{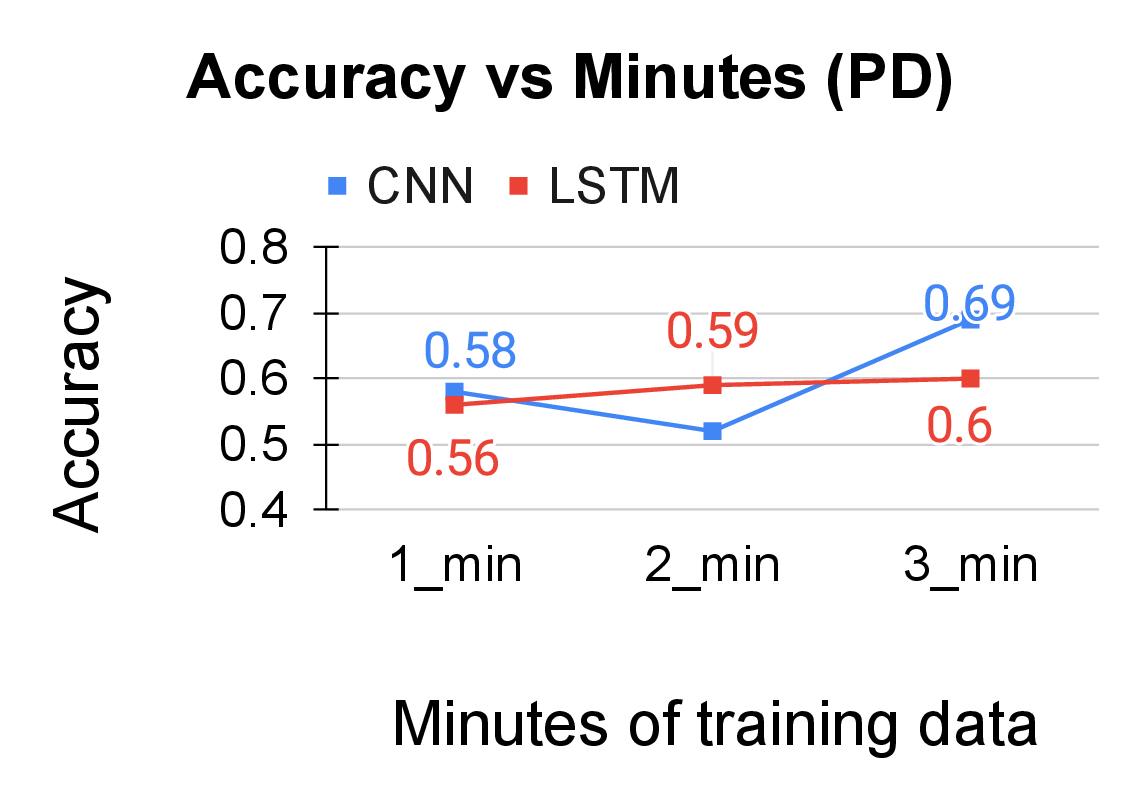} 
    \caption{CNN vs LSTM for PD}
    \label{fig:CNN_LSTM_PD}
  \end{subfigure}
  \hfill
  \begin{subfigure}{0.30\textwidth}
    \centering
    \includegraphics[width=\linewidth]{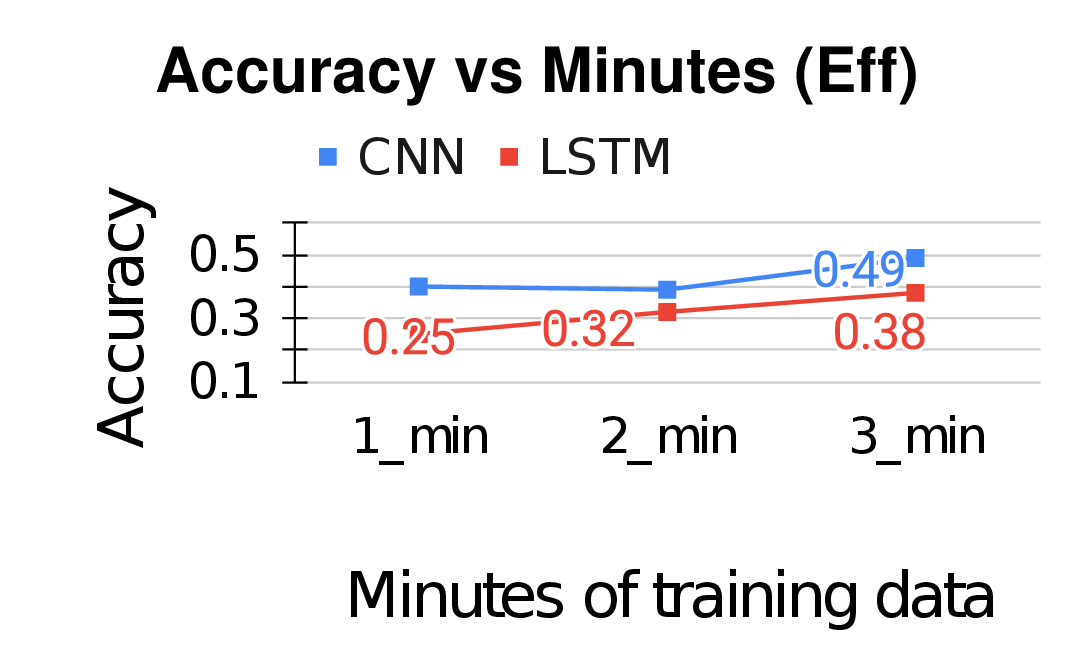} 
    \caption{CNN vs LSTM for Eff}
    \label{fig:CNN_LSTM_Eff}
  \end{subfigure}
  \begin{subfigure}{0.30\textwidth}
    \centering
    \includegraphics[width=\linewidth]{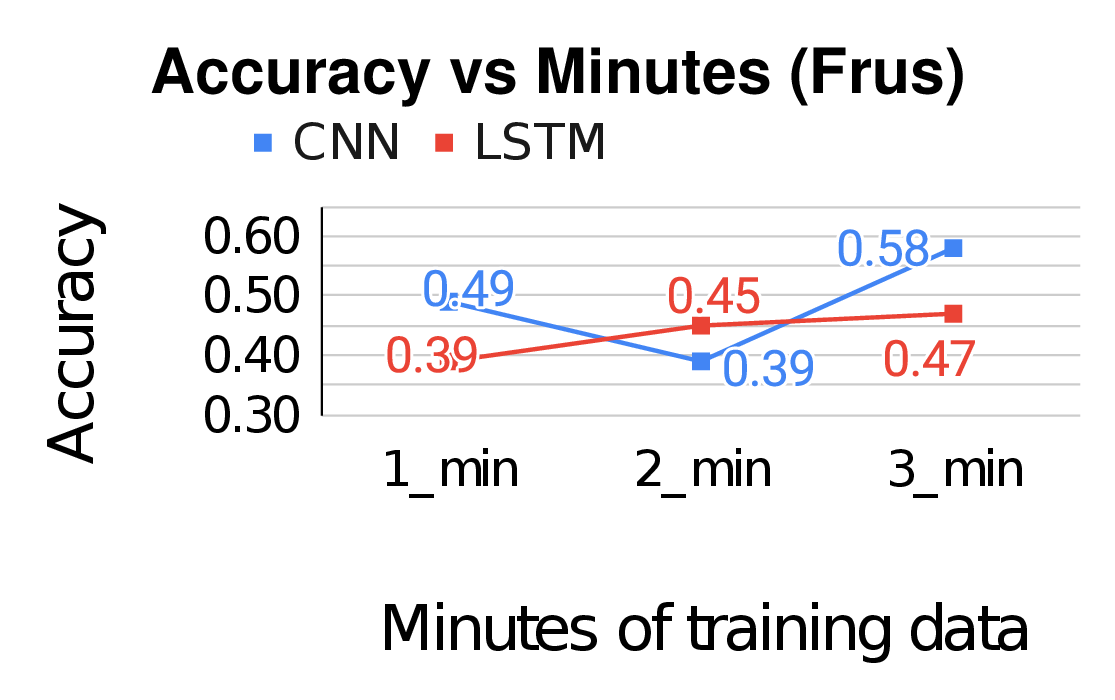} 
    \caption{CNN vs LSTM for Frus}
    \label{fig:CNN_LSTM_Frus}
  \end{subfigure}
  \caption{Comparing the performance of CNN and LSTM-based models for predicting responses on the NASA TLX. CNN outperforms LSTM in classification accuracies across all categories. }
  \label{fig:CNN_LSTM_NASA_TLX}
\end{figure*}
\section{Results and Discussions}\label{results_and_disc}
\subsection{Baseline Drift}
Since we had two baselines, at the beginning and end of the sessions, we checked the BVP and EDA data for baseline drifts using two sample t-tests based on the following hypothesis:
\begin{align}\label{eq:two_sample_t}
    \textrm{Null hypothesis (H}_{0}):  \mu_{B1}  =  \mu_{B2}\\
    \textrm{Alternate hypothesis (H}_{a}):  \mu_{B1} \neq \mu_{B2}
\end{align}
where $B1$ and $B2$ represent Baseline 1 and 2 respectively, for all the participants, the BVP data had no baseline drifts ($p-$value  $>0.05$, hence rejecting the null-hypothesis) but for the EDA sessions, we got $p-$value  $<0.05$. So we detrended all the sessions $k$ for the EDA data using scipy's signal processing library \cite{detrend} post which there was no baseline drift ($p-$value $>$ 0.05).

\subsection{Physiological difference between paper and robot-based instruction}\label{paper_vs_robot_physio}
We compared between using unimodal and multimodal data for both CNN and LSTM-based models. As shown in Figure \ref{fig:acc_vs_modal}, multimodal data outperforms unimodal data for both CNN and LSTM models. This can be attributed to the fact that using a multimodal pipeline (shown in Figure \ref{fig:model_pipelines}), helps learn from features extracted from both the BVP and EDA signals by concatenating the features before the classification layer. In addition to it, the CNN model outperforms the LSTM model by almost 7.72\%. The superior performance of the 1D CNN model is because it can capture short-range spatial features compared to the sequential nature of LSTMs \cite{hassanzadeh2024deep, marinho2023short}. 
\vspace{-0.5cm}
\subsection{Response analysis on NASA TLX}
We compared the differences in responses filled by the 22 participants on the NASA TLX between the printed and robot-based instruction tasks. The NASA TLX measures responses along six categories: mental demand (MD), temporal demand (TD), physical demand (PD), performance (Perf), effort (Eff), and frustration (Frus). For each category, we first checked for normality using the Ryan-Joiner test for normality using Minitab Statistical Software. For both the printed and robot-based tasks, responses for MD, PD, TD, Perf, and Eff were normally distributed ($p$-value $ > 1.0$). However, for Frus, the responses for the robot-based instruction task were normally distributed but not for the paper-based task ($p$-value  $= 0.016$). So, we used the paired t-test for comparing the printed and robot-based instruction tasks for the five categories except for Frus with the following hypothesis:
\begin{align}\label{eq:normality_test}
    \textrm{Null hypothesis (H}_{0}):  \mu_{p}  =  \mu_{r}\\
    \textrm{Alternate hypothesis (H}_{a}):  \mu_{p} \neq \mu_{r}
\end{align}
where, $\mu_{p}$ and $\mu_{r}$ represent the means of printed and robot-based instruction tasks respectively.
However, for comparing the responses for Frus, we used the Wilcoxon signed rank test based on the following hypothesis:
\begin{align}\label{eq:paired_t_test}
      \textrm{Null hypothesis (H}_{0}):  \eta = 0\\
    \textrm{Alternate hypothesis (H}_{a}): \eta \neq 0
\end{align}

where $\eta = \eta_{p} - \eta_{r}$, $\eta_{p}$, and $\eta_{r}$ represents the medians of the samples. The results for the paired t-test for MD, TD, PD, Perf, and Eff are shown in Figure \ref{fig:p_val_figure}. Between paper and robot-based instructions, temporal demand is significantly different ($p$-value  $= 0.013$), with the robot-based instruction having more temporal demand as compared to the paper-based task. This can be attributed to the fact that the participants could read the instructions for the printed-instruction based task at their own pace. In contrast, for the robot-based task, the instructions depended on how the robot delivered these instructions for assembling the Lego bricks. 

Further, Figure \ref{fig:wilcoxon} shows the Wilcoxon Signed Rank Test result. The median scores are not significantly different between the paper and robot-based instructions.

\subsection{Physiological difference based on workload}
In this paper, we investigated the use of physiological signals to predict the workload of individuals during printed and robot-based tasks. 
This helps eliminate the use of subjective measures in a longer study session to evaluate the participants' workload. So, we not only compare our CNN and LSTM-based models but also assess how early on in the session these pipelines could maximize their performance on the test data. Figure \ref{fig:CNN_LSTM_NASA_TLX} shows accuracy of the CNN and LSTM-based models for each category of NASA TLX (except for Perf). As seen from Figure \ref{fig:CNN_LSTM_NASA_TLX}, CNN-based models outperformed the LSTM based models in the first three minutes of training. The reason for choosing just three minutes is because some participants completed either the printed or robot-instruction based task in less than four minutes. Hence, we set the threshold for training data to three minutes. On an average, the CNN-based models performed approximately 17.83\% better than the LSTM-based model across MD, TD, PD, Eff, and Frus. The reason for leaving Perf out of training is that the percentage agreement between the reported performance on the NASA TLX and their actual performance was only 38.10\%. Hence, their reported scores on the task performance were not reliable. Here, the actual performance was evaluated to determine whether participants could follow all steps correctly or not.
\section{Conclusion}\label{conclusion}
In this paper, we associated collected physiological measures (BVP and EDA) from 22 participants with 1) a difference in printed and robot-based instruction setting, 2) a tool to predict the cognitive workload based on their response on the NASA TLX, both formulated as a classification problem. For 1), we compared our LSTM and CNN-based models using unimodal and multimodal data. The CNN-based approach outperformed the LSTM-based approach for classifying using multimodal data by 7.72\%. In addition to it, using multimodal data gave better results for classifying physiological measures between printed and robot-based instruction task (For CNN, approx. 8.38\% more for BVP and approx 20.49\% more for EDA and for LSTMs, 8.38\% more than just using BVP and 6.70\% more than just using EDA. For classifying responses on the NASA TLX using physiological signals, the same multimodal pipeline for CNN and LSTMs were compared. Our CNN-based model gave better classification results (approximately 17.83\% more on an average across all responses of the NASA TLX) within three minutes of training compared to the LSTM-based model.

\bibliographystyle{IEEEtran}
\bibliography{references}
\end{document}